# Small-signal Stability Analysis and Performance Evaluation of Microgrids under Distributed Control

Yimajian Yan, *Student Member, IEEE,* Di Shi, *Senior Member, IEEE,* Desong Bian, *Member, IEEE,* Bibin Huang, Zhehan Yi, *Member, IEEE,* Zhiwei Wang, *Senior Member, IEEE*

***Abstract*—Distributed control, as a potential solution to decreasing communication demands in microgrids, has drawn much attention in recent years. Advantages of distributed control have been extensively discussed, while its impacts on microgrid performance and stability, especially in the case of communication latency, have not been explicitly studied or fully understood yet. This paper addresses this gap by proposing a generalized theoretical framework for small-signal stability analysis and performance evaluation for microgrids using distributed control. The proposed framework synthesizes generator and load frequency-domain characteristics, primary and secondary control loops, as well as the communication latency into a frequency-domain representation which is further evaluated by the generalized Nyquist theorem. In addition, various parameters and their impacts on microgrid dynamic performance are investigated and summarized into guidelines to help better design the system. Case studies demonstrate the effectiveness of the proposed approach.**

***Keywords*—Microgrids, distributed control, communication delay, small-signal stability, generalized Nyquist stability criterion.**

## NOMENCLATURE

| | |
|---|---|
| $p$ | time derivative operator, $s^{-1}$ |
| **G** | directed graph |
| **N** | node set of a directed graph |
| **E** | edge set of a directed graph |
| **A** | adjacent matrix |
| **D** | degree matrix |
| $n$ | number of DGs in a system |
| $N_i$ | neighbor set of node $i$ |
| $d_i$ | In-degree coefficient of the $i^{th}$ DG |
| **L** | Laplacian matrix |
| $\boldsymbol{\theta}$ | input vector |
| $\theta_i$ | the $i^{th}$ element in an input vector |
| $\tau_{ij}$ | latency from node $j$ to node $i$, s |
| $\omega_i^*$ | desired output frequency of the $i^{th}$ DG, rad/s |
| $\omega_{ref}$ | input reference frequency |
| $V_{ref}$ | input reference RMS voltage |
| $\delta_i$ | angle of the $i^{th}$ DG with respect to slack bus, rad |
| $V_i^*$ | desired output voltage of the $i^{th}$ DG, V |
| $k_{pi}$ | frequency droop coefficient of the $i^{th}$ DG, rad/(s·W) |
| $k_{qi}$ | voltage droop coefficient of the $i^{th}$ DG, V/VAR |
| $\hat{f}$ | latened function/variable $f$ |
| $f_d$ | direct component of $f$ |
| $f_q$ | quadrant component of $f$ |
| $P_i$ | active power of the $i^{th}$ DG, W |
| $Q_i$ | reactive power of the $i^{th}$ DG, VAR |
| $J$ | Normalized inertia of a generator, kg·m$^2$ |
| $\omega_b$ | base (nominal) frequency of generators, rad/s |
| $k_1$ | proportional gain of PI control |
| $k_2$ | integral gain of PI control |
| $\sigma$ | power output filter time constant |
| $\sigma_v$ | voltage low-pass filter time constant |
| $D$ | damping torque factor, N·m·s |
| $D_p$ | damping factor ($D_p = D\cdot\omega_b$), N·m |

This work is supported by SGCC Science and Technology Program under project Distributed Fast Frequency Control under HVDC Line Faulty Conditions.

Y. Yan, D. Shi, D. Bian, Z. Yi, and Z. Wang are with GEIRI North America, San Jose, CA 95134. Email: yan143@purdue.edu; di.shi@geirina.net.

B. Huang is with State Grid Energy Research Institute, Beijing, China. Email: huangbibin@sgeri.sgcc.com.cn.

## I. INTRODUCTION

As penetration levels of distributed energy resources (DERs) keep increasing, the centralized control paradigm for microgrid is facing great challenges from enormous communication demands. To address these challenges, a number of algorithms have been proposed in recent years, such as distributed control, inverter-based autonomous operation [1], etc. Among them, the ones based on the consensus concept have received much attention. Authors in [1] and [3] proposed consensus-based distributed primary and secondary controls for microgrids. Authors of [4]-[6] implemented consensus-based distributed tertiary control to achieve optimal economic dispatch for microgrids.

Although distributed control brings advantages, one major concern for microgrids under distributed control is system dynamic performance and stability, which becomes a complicated issue due to the communication latency and its underlying uncertainty. Instability undermines system resilience and should be avoided at the design stage. Indeed, small-signal stability analysis for microgrids under centralized hierarchical controls has been extensively studied [6]-[8], but little work has been done investigating performance of microgrids under distributed control, especially with communication latency considered.

Convergence of consensus-based tertiary control is investigated in [9] and [10] through discrete-time systems analysis. The conclusions may not hold if communication delay is considered. Authors in [11] show that communication delay has an impact on the resilience of microgrid under distributed control with the shortage of theories. Reference [12] studies microgrid stability through eigenvalue analysis, which unfortunately is not eligible for microgrids under distributed control with communication latency. Papers [13]-[16] investigate the small-signal stability of microgrids with distributed control by solving delay differential equations (DDEs), which is hard to represent system performance without gain and phase margins, and difficult to address different latencies in different communication edges. Authors of [17] derive a latency threshold using the Lyapunov-Krasovkii function but only consider frequency regulation. Authors of [18] discuss latency thresholds for microgrids under consensus-based distributed control, but only for strongly-connected graphs and inverter-based distributed generators (DGs) (in which the rotor inertia is not involved). In [19], the authors proposed a microgrid networked control scheme that is robust to communication latency and even data losses. Nonetheless, neither proof nor detailed analysis was provided. Besides, there have been some discussions on the stability of time-delay systems [20]-[22] in general automatic control areas, but they have not been implemented to microgrids. In summary, existing works has not proposed a methodology to comprehensively study all the factors of distributed control of microgrids.

This paper addresses the gap by making contributions from the following three aspects. First, an innovative approach is proposed for small-signal stability analysis for microgrids under distributed control with communication latency and uncertainty considered. Second, a generalized model is presented which regards any microgrid component, linear or nonlinear, as a frequency-domain transfer function or a matrix of transfer functions. Generalized Nyquist theorem for multiple input multiple output (MIMO) systems is applied to this model, which not only identifies stability but also the dynamic performance of the system. Third, various parameters and their impacts on microgrid dynamic performance are investigated using the proposed framework, which provides a reference for generating guidelines in practice.

The remainder of this paper is organized as follows. Section II revisits graph theory, distributed control algorithms, and the model for communication latency. Section III presents the proposed frequency-domain model of microgrid and the proposed approach for microgrid stability and performance assessment, using generalized Nyquist criterion for multiple-input-multiple-output (MIMO) system. Section IV introduces an exemplary four-bus microgrid model and presents six case studies that validate the proposed framework, together with detailed analysis of various design parameters and their impacts on the dynamic performance of microgrids. Conclusions and future work are discussed in Section V.

## II. Preliminaries

### A. Graph Theory

Suppose there are $n$ nodes/agents in a network. The corresponding communication topology can be represented by a directed graph $G=\{N, E\}$, where $N=\{1,\ldots,n\}$ is the node set and $E$ is the edge set. Each edge $(i, j)$ represents a path of information flow from node $j$ to node $i$. Neighbors of node $i$ are defined as $N_i=\{j\in N: (i, j)\in E\}$. According to this definition, node $i$ only has access to the information from its neighbors defined by $N_i$.

Further, an adjacency matrix can be defined as $\mathbf{A}=[a_{ij}]\in\mathbb{R}^{n\times n}$ (note that a bold symbol represents a vector of matrix), where $a_{ij}=1$ if $j\in N_i$, and $a_{ij}=0$ otherwise. An in-degree matrix is defined as $\mathbf{D}=[d_{ij}]\in\mathbb{R}^{n\times n}$ where $d_{ij}=0$ for $i\neq j$, and $d_{ii}=d_i=\sum_j a_{ij}$. The Laplacian matrix of the directed graph can be defined as $\mathbf{L}=\mathbf{D}-\mathbf{A}$.

For a first-order multi-agent system, let node $i$ has a single state $x_i$ so that its time-domain dynamics can be described as:

$$px_i=\sum_{j\in N_i} a_{ij}\left(x_j-x_i\right) \tag{1}$$

In (1) and the following equations, $p$=d/d$t$ is the time differential operator. Define the system state vector as $\mathbf{x}=[x_1,\ldots, x_n]^T$. Then time-domain dynamics of the system using consensus-based control based on graph $G$ is governed by:

$$p\mathbf{x}=-\mathbf{L}\mathbf{x} \tag{2}$$

The nodes in the graph that receive an external reference signal $u_{ref}$ are defined as leading nodes. An input vector $\boldsymbol{\theta}=[\theta_1,\ldots\theta_n]^T$ can be defined as $\theta_i=1$ if node $i$ is a leading node, and $\theta_i=0$ otherwise. Therefore, dynamics of the first-order system with external reference signal can be expressed as:

$$p\mathbf{x}=-\mathbf{L}\mathbf{x}+u_{ref}\boldsymbol{\theta}-\mathrm{diag}(\boldsymbol{\theta})\mathbf{x} \tag{3}$$

Communication networks for a microgrid under distributed control can have very flexible topology, as long as some basic rules are followed [3]. If each node in a directed graph retains a directed path to every other node, this graph is strongly connected. In contrast, a special case of a weakly-connected directed graph is a directed spanning tree. It has one leading node called "root node", which has a directed path to all other nodes [20]. Fig. 1 presents a strongly-connected directed graph and a directed spanning tree.

It has been shown in [3] and [20] that if a consensus-based system includes at least one leading node and a directed spanning tree from the leading node(s), all $x$'s will converge to the external reference $u_{ref}$. In a microgrid where multiple control loops exist, responses of $x$'s to their neighboring states are complicated, which will be further discussed in section III.

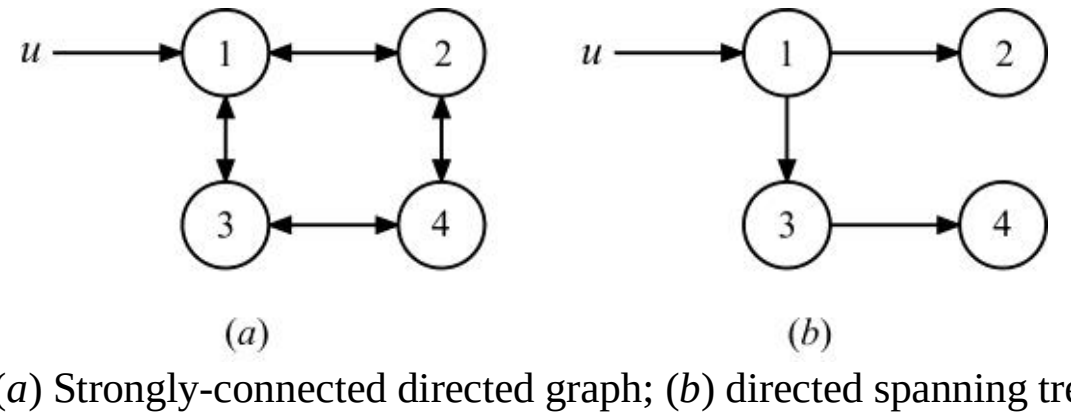

Fig. 1 ($a$) Strongly-connected directed graph; ($b$) directed spanning tree

### B. Distributed Control with Communication Delay

Different variants of the control law described by (3) can be used to achieve distributed control for microgrids. To name a few, references [1]-[3] propose distributed droop control schemes to realize voltage and frequency regulation while making sure the power-sharing among DGs follows pre-defined rules; [4]-[6] propose distributed tertiary control to realize optimal economic dispatch. This paper mainly focuses on the droop and secondary control loops as dynamic performance of microgrids are largely dependent upon them. The distributed controls presented in [1]-[3] are used for discussion in this work while the proposed approach can be easily extended to accommodate other variants.

Starting from the droop control, the droop characteristic of a DG can be expressed as

$$\omega_i = \omega_i^* - k_{pi} P_i \tag{4}$$

$$V_i = V_i^* - k_{qi} Q_i \tag{5}$$

The droop coefficients $k_{pi}$'s and $k_{qi}$'s are typically set inversely proportional to capacities of DGs for power sharing, although there can be many different ways [11]. In steady state, equalization of both $k_{pi} P_i = k_{pj} P_j$ and $\omega_i = \omega_j$ for $i, j \in N$ can be reached since the transmission of active power between buses or DGs is dependent upon phase angle differences, i.e., the time integration of output frequency. To implement this control law in a microgrid, $DG_j$ ought to spread its value of $\omega_j + k_{pj} P_j$ to the communication network. The receiver $DG_i$, when $a_{ij}=1$, will receive $\widehat{\omega}_j + k_{pj} \widehat{P}_j$ due to the latency, and update its own $\omega_i$ accordingly. Herein, a signal $f$ with latency is labeled as $\widehat{f}$, and since the droop coefficients $k_p$'s are constants, they are not so labeled. Additionally, if $DG_i$ is also attached to the input references $\omega_{ref}$ and $V_{ref}$, it is also forced to follow these references. Therefore, the first law of consensus-based distributed control of $DG_i$ can be formulated as (6):

$$p\omega_i^* = \sum_{j \in N_i} a_{ij} \left( \widehat{\omega}_j - \omega_i + k_{pj} \widehat{P}_j - k_{pi} P_i \right) + \theta_i \left( \omega_{ref} - \omega_i \right) \tag{6}$$

For voltage control, accurate reactive power sharing is difficult to achieve with (5) due to transmission line impedance variations. Therefore, [3] proposed a control method so that each DG's output voltage can be fully controlled by achieving a consensus on voltage droop displacement $k_{qi} Q_i$. Although this method realized accurate reactive power (proportional to capacity) sharing between DGs, the voltage differences can be substantial if the line is long (with significate impedance). In this research, the sharing of reactive power is not the primary objective of the control. For simplicity, assign small $k_{qi}$ values and include $V_i^* = V_j^*$ for $i, j \in N$ in the consensus protocol. Accordingly, the second law of consensus-based distributed control of $DG_i$ (also considering following the reference for leading nodes) can be written as:

$$pV_i^* = \sum_{j \in N_i} a_{ij} \left( \widehat{V}_j - V_i + k_{qj} \widehat{Q}_j - k_{qi} Q_i \right) + \theta_i \left( V_{ref} - V_i \right) \tag{7}$$

Eventually, with (7), the DG's in a microgrid will output with slightly different voltages, in order to reach a consensus on the values of $V + k_q Q$ through the communication network. The advantage of the introduced consensus protocol, including the two control laws (6) and (7), is that only two variables ($\omega + k_p P$ and $V + k_q Q$) are needed to be sent through the communication network in one direction. This allows a low-cost communication infrastructure to be utilized for a smart grid without sacrificing reliability and redundancy.

## III. Proposed Model and Approach

This section presents a generalized state-space model for microgrids under distributed control with communication latency considered and the corresponding approach for microgrid dynamic performance evaluation. In this section, if not specified, all variables are evaluated in frequency-domain ($s$-domain with $s=j\omega$).

### A. Proposed Microgrid Model

For small-signal stability analysis, the system state vector of an $n$-bus microgrid can be defined as

$$\mathbf{x} = \left[ \Delta\omega_1, \Delta\omega_2, ..., \Delta\omega_n, \Delta V_1, \Delta V_2, ..., \Delta V_n \right]^T \tag{8}$$

where the small-signal of the output frequencies and RMS voltages are included. Meanwhile, a DG power output vector is defined as

$$\mathbf{y} = \left[ \Delta P_1, \Delta P_2, ..., \Delta P_n, \Delta Q_1, \Delta Q_2, ..., \Delta Q_n \right]^T \tag{9}$$

Both **x** and **y** are evaluated around some equilibrium point. The equilibrium point can be obtained by solving $\omega$'s and $V$'s using $\omega_i = \omega_j$ and $V_i + k_{qi} Q_i = V_j + k_{qj} Q_j$ for all DGs. This is carried out using population-based optimization in the case studies presented in the next section.

For a continuous signal $f(t)$, the corresponding delayed signal can be defined as $f(t-\tau)$, where $\tau$ is the time latency. This latency can also be a continuous signal $\tau(t)$ which varies with time. The frequency-domain representation of this is shown in (10):

$$\mathcal{L}\left\{ f\left(t-\tau\right) \right\} = e^{-\tau s} F\left(s\right) \tag{10}$$

Using this equation, the delay operation can be converted to a transfer function and be treated identically with another type of transfer functions such as a low-pass filter.

Suppose that each edge $(i, j) \in E$ has a latency $\tau_{ij} \geq 0$. The time-domain delay differential equations (6) and (7) for all DGs, using the definition in (8) and (9), can be transformed to frequency-domain as

$$s\mathbf{x} - \mathbf{x}_0 = \left( \begin{bmatrix} \widehat{\mathbf{A}} - \mathbf{D} & \mathbf{0} \\ \mathbf{0} & \widehat{\mathbf{A}} - \mathbf{D} \end{bmatrix} \right) \left( \mathbf{x} + \mathbf{K}\mathbf{y} \right) \tag{11}$$

where **D** is the degree matrix as defined in subsection II. $A$, $\mathbf{x}_0$ is the initial value of **x**, and

$$\widehat{\mathbf{A}} = \left[ e^{-\tau_{ij} s} a_{ij} \right], \text{ for } i, j \in N \tag{12}$$

$$\mathbf{K} = \begin{bmatrix} \mathbf{K}_p & \mathbf{0} \\ \mathbf{0} & \mathbf{K}_q \end{bmatrix} = \text{diag}\left( \left[ k_{p1}, ..., k_{pn}, k_{q1}, ..., k_{qn} \right] \right) \tag{13}$$

The frequency-domain characteristics described in (11) only consider the distributed secondary control of DGs. The inner control loops, filters and rotor (virtual) inertia should also be included. For instance, frequency control of a DG may employ a PID controller to regulate its mechanical power, and the slew rate of the speed of a rotor is influenced by rotor inertia. Generally, in a similar form to (11), the following expression illustrates the system characteristics:

$$s\mathbf{x} - \mathbf{x}_0 = \mathbf{A}_s \mathbf{x} + \mathbf{B}_s \mathbf{y} \tag{14}$$

where both $\mathbf{A}_s$ and $\mathbf{B}_s$ are $2n$-by-$2n$ matrices consisting of $s$-domain functions. Basically, there are two approaches to obtain matrix $\mathbf{A}_s$ and $\mathbf{B}_s$: from a design perspective, the control and characteristics of each DG can be modeled in $s$-domain as transfer functions, as to be shown in an example in the next section; from a planning perspective, the frequency-domain characteristics of a component can be measured using small-signal sinusoidal injection [19].

It can also be found that $\mathbf{y}=\mathbf{C}_s\mathbf{x}$ where $\mathbf{C}_s$ is a $2n$-by-$2n$ matrix of $s$-domain functions. To derive $\mathbf{C}_s$, the system state vector $\mathbf{x}$ is first converted to $dq$ reference frame. Towards this end, the direct and quadrant values of a phasor $\tilde{f} = F\angle\delta_f$ can be defined as

$$f_d = F\cos\delta_f \quad (15)$$

$$f_q = F\sin\delta_f \quad (16)$$

For the $i$th DG, define the following functions:

$$m_{dei} = \frac{v_{qei}}{v_{dei}^2 + v_{qei}^2},\; m_{qei} = \frac{v_{dei}}{v_{dei}^2 + v_{qei}^2} \quad (17)\text{-}(18)$$

$$n_{dei} = \frac{v_{dei}}{\sqrt{v_{dei}^2 + v_{qei}^2}},\; n_{qei} = \frac{v_{qei}}{\sqrt{v_{dei}^2 + v_{qei}^2}} \quad (19)\text{-}(20)$$

$$D_{ei} = m_{dei}n_{qei} - n_{dei}m_{qei} \quad (21)$$

$$\mathbf{M}_{ei} = \frac{1}{D_{ei}}\begin{bmatrix} n_{qei} & -m_{qei} \\ -n_{dei} & m_{dei} \end{bmatrix} \quad (22)$$

where $v_{dei}$ and $v_{qei}$ are the $d$-axis and $q$-axis components of a voltage at the system equilibrium point, respectively. According to these definitions, a state vector made up of DGs' terminal voltages in the $dq$ reference frame can be written as:

$$\Delta\mathbf{v}_{dq} = \left[\Delta v_{d1}, \Delta v_{q1}, ..., \Delta v_{dn}, \Delta v_{qn}\right]^T = \mathbf{MTEx} \quad (23)$$

$$\mathbf{M} = \begin{bmatrix} \mathbf{M}_{e1} & & \\ & \ddots & \\ & & \mathbf{M}_{en} \end{bmatrix}, \text{ and } \mathbf{E} = \begin{bmatrix} \frac{1}{s}\mathbf{I}_n & \\ & \mathbf{I}_n \end{bmatrix} \quad (24)\text{-}(25)$$

where $\mathbf{T}$ is an orthogonal matrix used to rearrange elements in $\mathbf{Ex}=[\delta_1, \ldots,\delta_n, V_1, \ldots, V_n]^T$ so that $\mathbf{TEx}=[\delta_1, V_1, \ldots,\delta_n, V_n]^T$.

With voltage and current of $\text{DG}_i$ known, its active and reactive power outputs, which are elements of $\mathbf{y}$, can be calculated as:

$$P_i = 3v_{di}i_{di} + 3v_{qi}i_{qi},\; Q_i = 3v_{qi}i_{di} - 3v_{di}i_{qi} \quad (26)\text{-}(27)$$

In small-signal analysis, combining (26) and (27) yields

$$\begin{bmatrix} \Delta P_i \\ \Delta Q_i \end{bmatrix} = 3\mathbf{i}_{dqei}\Delta\mathbf{v}_{dqi} + 3\mathbf{v}_{dqei}\Delta\mathbf{i}_{dqi} \quad (28)$$

where

$$\mathbf{i}_{dqei} = \begin{bmatrix} i_{dei} & i_{qei} \\ -i_{qei} & i_{dei} \end{bmatrix}, \text{ and } \mathbf{v}_{dqei} = \begin{bmatrix} v_{dei} & v_{qei} \\ v_{qei} & -v_{dei} \end{bmatrix} (29)\text{-}(30)$$

Denoting system admittance matrix as $\mathbf{Y}_s$, (28) can be simplified by substituting $\Delta\mathbf{i}_{dq}=\mathbf{Y}_s\Delta\mathbf{v}_{dq}$, which leads to:

$$\mathbf{Ty} = \left[\Delta P_1, \Delta Q_1, \ldots, \Delta P_n, \Delta Q_n\right]^T = 3\left(\mathbf{i}_{dqe} + \mathbf{v}_{dqe}\mathbf{Y}_s\right)\Delta\mathbf{v}_{dq} \quad (31)$$

where

$$\mathbf{i}_{dqe} = \text{diag}\left(\left[\mathbf{i}_{dqe1}, ..., \mathbf{i}_{dqen}\right]\right) \quad (32)$$

and

$$\mathbf{v}_{dqe} = \text{diag}\left(\left[\mathbf{v}_{dqe1}, ..., \mathbf{v}_{dqen}\right]\right) \quad (33)$$

Combining (23) and (31) gives $\mathbf{C}_s$ as

$$\mathbf{C}_s = \mathbf{T}^{-1}\left(\mathbf{i}_{dqe} + \mathbf{v}_{dqe}\mathbf{Y}_s\right)\mathbf{MTE} \quad (34)$$

In practice, real-time measurements are first passed through a low-pass filter before being used to calculate the power output of a DG. Therefore, a factor of $1/(\sigma s+1)$ should be applied to (34), where $\sigma$ is the time-constant of the low-pass filter.

To implement (14) in a microgrid under distributed control, $\mathbf{A}_s$, $\mathbf{B}_s$, and $\mathbf{Y}_s$ should be evaluated based on system parameters and setups. In general, rewriting (14) for a microgrid model yields

$$\mathbf{x} = \left[\mathbf{I}_{2n} - \frac{1}{s}\left(\mathbf{A}_s + \mathbf{B}_s\mathbf{C}_s\right)\right]^{-1}\frac{1}{s}\mathbf{x}_0 \quad (35)$$

which can be regarded as a transfer function of a closed-loop system with input $\mathbf{x}_0$, output $\mathbf{x}$, and a return function $\mathbf{L}(s) = -(\mathbf{A}_s + \mathbf{B}_s\mathbf{C}_s)/s$.

### *B. Proposed Dynamic Performance Evaluation Approach*

Nyquist stability criterion is a widely used graphic technique for determining the stability of a dynamic system. The advantage of Nyquist-based stability analysis is that the frequency-domain system state-space realizations $\{\mathbf{A}(s), \mathbf{B}(s), \mathbf{C}(s), \mathbf{D}(s)\}$ (whose eigenvalues are no longer fixed points in the complex plane, but a function of $s$ instead) can be directly utilized. This greatly simplifies mathematical analysis for high-order or nonlinear systems.

The generalized Nyquist theorem for multiple input multiple output (MIMO) systems is stated in the following theorem [24] with an illustration in Fig. 2, where the input $\mathbf{u}$ and output $\mathbf{y}$ are vectors, and $\mathbf{G}(s)$ and $\mathbf{K}(s)$ are matrices of transfer functions. The full mathematical proof of the generalized Nyquist criterion is given in [24].

**Theorem**: Let the MIMO system in Fig. 2 have no open-loop uncontrollable or unobservable modes whose corresponding characteristic frequencies lie in the right-half plane. Then this system will be closed-loop stable if and only if the sum of counter-clockwise encirclements around the critical point ($-1+j0$) by the set of characteristic loci of the return ratio $\mathbf{L}(s) = \mathbf{G}(s)\mathbf{K}(s)$ is equal to the total number of right-half plane poles of $\mathbf{G}(s)$ and $\mathbf{K}(s)$. The characteristic loci are the set of positions of eigenvalues of $\mathbf{L}(s)$ on the complex plane evaluated for all value of $s$ on the Nyquist contour.

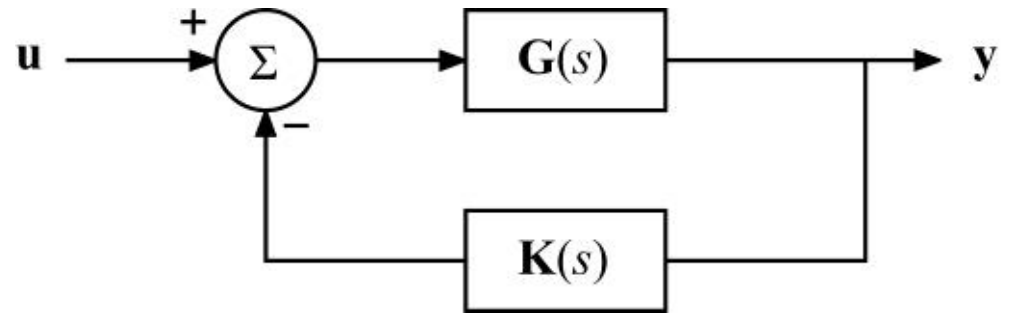

Fig. 2 MIMO feedback system

Using (35), a microgrid working under distributed control can be modeled as shown in Fig. 2 with $\mathbf{u} = \mathbf{x}_0$, $\mathbf{y} = \mathbf{x}$, $\mathbf{G}(s) = \mathbf{I}_{2n}/s$, and $\mathbf{K}(s) = -(\mathbf{A}_s + \mathbf{B}_s\mathbf{C}_s)$. From derivations of $\mathbf{A}_s$, $\mathbf{B}_s$ and $\mathbf{C}_s$ above, it is easy to show that $\mathbf{G}(s)$ and $\mathbf{K}(s)$ will not have any right-half plane poles. Thus, the system will be stable if and only if the characteristic loci of the return ratio $\mathbf{L}(s) = -(\mathbf{A}_s+\mathbf{B}_s\mathbf{C}_s) / s$ does not encircle ($-1+j0$).

In addition, the system dynamic performance can be evaluated by analyzing gain margin (GM) and phase margin (PM) of the Nyquist plot. A system with larger GM and PM is more robust to parameter errors and changes. GM can also be used to evaluate the rate of convergence when the system is stable, since it is equal to the absolute value of the largest non-zero eigenvalue of the closed-loop system [25]. Note that for multiple characteristic loci rather than one locus, the smallest GM and PM should be considered. If GM<1 or PM<0, the encirclement of the critical point (-1+*j*0) will occur and the system becomes unstable.

## IV. Microgrid Model and Case Studies

### *A. An Exemplary Analysis of a Four-bus Microgrid*

An islanded four-bus microgrid shown in Fig. 3 is selected as an exemplary system to illustrate the proposed methodology. As Fig. 3 shows, each bus is connected to DG and load. Bus 1 is set as PCC (point of common coupling) and its voltage angle is set to zero (slack bus). All transmission line and load impedances are evaluated as transfer functions Y(*s*)'s. For simplicity, suppose that the active and reactive power load on a bus is given and the load only contains passive elements (R's, L's and C's). Therefore, the load impedance Y(*s*) can be calculated simply from the given load active and reactive power. In further researches, active load characteristics (e.g. constant current load or constant power load) can be easily introduced to this framework by replacing a load impedance Y(s) with a more complicated frequency-domain function.

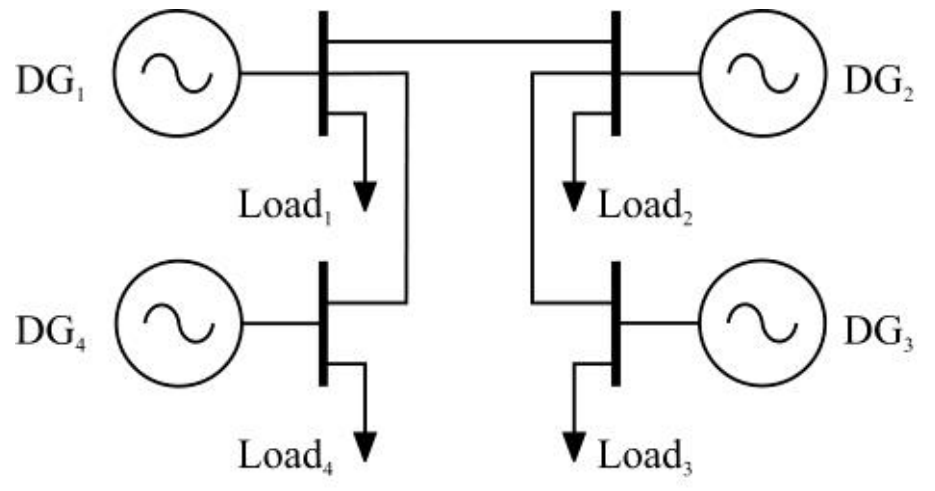

Fig. 3 An exemplary 4-bus microgrid

In a simplified form, the control of each DG is shown in Fig. 4. In detail, Fig. 4(*a*) shows the distributed frequency control loop for DG$_i$. For active power sharing between DGs, the average of all neighbors' frequencies, the microgrid frequency reference, and the droop component are synthesized. For the $i^{\text{th}}$ DG control, the acquired frequency from its neighbor *j* is delayed for $\tau_{ij}$, which is shown in Fig. 4 as a transfer function. Consequently, the following equation can be formulated:

$$\omega_i^* = \frac{1}{d_i+\theta_i}\left(\sum\nolimits_{j\in N_i} e^{-\tau_{ij}s} a_{ij}\omega_j + \theta_i \omega_{ref}\right) + \sum\nolimits_{j\in N_i}\left(e^{-\tau_{ij}s} a_{ij} k_{pj} P_j - k_{pi} P_i\right). \tag{36}$$

Fig. 4(b) shows the distributed voltage regulation control diagram for DG$_i$. The LPF (low-pass filter) in Fig. 4(c) provides a simplified model (which can be further addressed with a more realistic model) of the combined internal winding inductance, capacitance and parasitic paths between the generator and the bus. The following equation can be derived:

$$sV_i^* = \sum\nolimits_{j\in N_i}\left[e^{-\tau_{ij}s} a_{ij}\left(V_j + k_{qj}Q_j\right) - V_i - k_{qi}Q_i\right] + \theta_i\left(V_{ref} - V_i\right) \tag{37}$$

Nowadays, inverter-based DGs, which utilize renewable energies or storage units, are being increasingly designed as virtual synchronous generators (VSG) [26, 27] to provide damping effects. These inverter-based DGs behave very similarly to synchronous generators (SG). For illustration purpose, Fig. 4(*c*) shows a simplified representative model of such DG. From this figure, the dynamics of the output frequency of DG$_i$ are governed by:

$$s\omega_i = \frac{1}{J\omega_b}\left[\left(k_1 + \frac{k_2}{s}\right)\left(\omega_i^* - \omega_i\right) - P_i - D_p\left(\omega_i - \omega_b\right)\right] \tag{38}$$

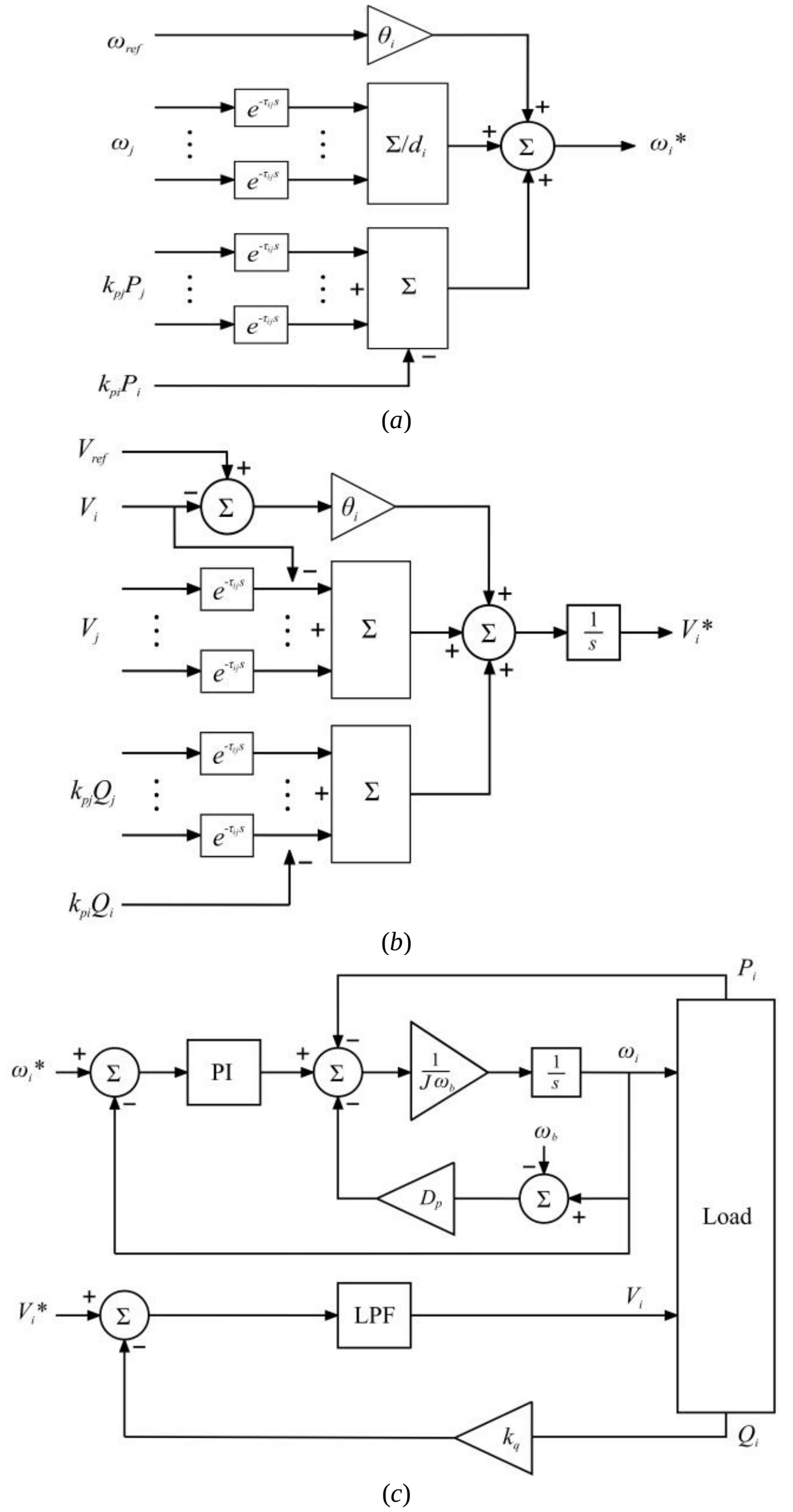

Fig. 4 (a) frequency consensus-based control, (b) voltage consensus-based control, (c) DG model.

The frequency transfer function of DG$_i$ can be written as:

$$\omega_i = T_\omega \omega_i^* - T_p P_i \tag{39}$$

where

$$T_\omega = \frac{k_1 s + k_2}{J\omega_b s^2 + \left(k_1 + D\right)s + k_2}$$

$$T_p = \frac{s}{J\omega_b s^2 + \left(k_1 + D\right)s + k_2}$$

Meanwhile, the voltage transfer function of DG$_i$ can be written as:

$$V_i = T_v\left(V_i^* - k_{qi} Q_i\right) \tag{40}$$

where $T_v$ is the transfer function of the LPF which is defined as $T_v$=1/($\sigma_v s$+1).

As inputs $\omega_{ref}$ and $V_{ref}$ can be regarded as constants (for a long enough time in practice, but can change for synchronized reconnection [11]), they typically have no impact on system small signal analysis. Therefore, according to (14), combining the aforementioned transfer functions yields

$$\mathbf{A}_s = \begin{bmatrix} sT_\omega\left(\mathbf{D} + \mathrm{diag}(\boldsymbol{\theta})\right)^{-1}\widehat{\mathbf{A}} & \mathbf{0} \\ \mathbf{0} & T_v\left(\widehat{\mathbf{A}} - \mathbf{D} - \mathrm{diag}(\boldsymbol{\theta})\right) \end{bmatrix} \tag{41}$$

$$\mathbf{B}_s = \begin{bmatrix} sT_\omega\left(\widehat{\mathbf{A}} - \mathbf{D}\right)\mathbf{K}_q - sT_p\mathbf{I}_n & \mathbf{0} \\ \mathbf{0} & T_v\left(\widehat{\mathbf{A}} - \mathbf{D} - s\mathbf{I}_n\right)\mathbf{K}_q \end{bmatrix} \tag{42}$$

where $\widehat{\mathbf{A}}$ is defined in (12), $\mathbf{D}$ is the in-degree matrix, $\mathbf{K}_p$ and $\mathbf{K}_q$ are defined in (13).

To derive the system admittance matrix (for *dq* decomposed analysis) $\mathbf{Y}_s$, start from the system topology in Fig. 3 with the phasor analysis. The relationship between the output current phasors and the output voltage phasors of all DGs can be expressed as $\tilde{\mathbf{I}} = \mathbf{Y}\tilde{\mathbf{V}}$ where

$$\mathbf{Y} = \begin{bmatrix} Y_1 + Y_{12} + Y_{14} & -Y_{12} & 0 & -Y_{14} \\ -Y_{12} & Y_2 + Y_{12} + Y_{23} & -Y_{23} & 0 \\ 0 & -Y_{23} & Y_3 + Y_{23} & 0 \\ -Y_{14} & 0 & 0 & Y_4 + Y_{14} \end{bmatrix} \tag{43}$$

From (15) and (16), a phasor can be obtained from its corresponding *dq* values according to

$$\tilde{f} = f_d + jf_q \tag{44}$$

Therefore, considering Y =G + jB, (45) can be derived:

$$\mathbf{Y}_s = \begin{bmatrix} \mathbf{Y}_{dq1} + \mathbf{Y}_{dq12} + \mathbf{Y}_{dq14} & -\mathbf{Y}_{dq12} & 0 & -\mathbf{Y}_{dq14} \\ -\mathbf{Y}_{dq12} & \mathbf{Y}_{dq2} + \mathbf{Y}_{dq12} + \mathbf{Y}_{dq23} & -\mathbf{Y}_{dq23} & 0 \\ 0 & -\mathbf{Y}_{dq23} & \mathbf{Y}_{dq3} + \mathbf{Y}_{dq23} & 0 \\ -\mathbf{Y}_{dq14} & 0 & 0 & \mathbf{Y}_{dq4} + \mathbf{Y}_{dq14} \end{bmatrix} \tag{45}$$

where

$$\mathbf{Y}_{dqx} = \begin{bmatrix} G_x & -B_x \\ B_x & G_x \end{bmatrix} \tag{46}$$

for $x$= 1, 12, 14, …, etc.

In summary, $\mathbf{A}_s$, $\mathbf{B}_s$, and $\mathbf{Y}_s$ have been derived and the system small-signal model (35) has been realized for this exemplary 4-bus microgrid. Besides, although this paper only considers a simplified model of the DGs (similar but more representative compared to previous works [2-5]), more complicated systems can be represented as well using (35) with detailed models of components. Further, this framework also works for inverter-based DGs with known state-space models, which can be easily converted to *s*-domain functions as (39) and (40).

Parameters of the 4-bus microgrid are listed in Table I and II. The latency in the microgrid network can be addressed in following aspects: (1) latency from system status change and the response of the measurement, which may vary from 0.035 to 1.038s according to IEEE Standard C37.118.1™ if PMU serves as the measuring device; (2) latency from data processing, which may occurs on both sender and receiver sides, and may include algorithm processing, data compression, data encryption, etc., which usually costs 0.1 to 0.3s [28]; (3) latency from information traveling between the sender and the receiver within communication networks which is around 0.3s normally [29]. In conclusion, a reasonable approximation of the total latency in the secondary control is from 0.5s to 2.5s [30]. Depending upon system parameters and designs, it is safe to conclude it is possible for a microgrid under distributed control to be unstable even with small latencies.

TABLE I SYSTEM PARAMETERS

| **Transmission Line Parameters** | | | | **Load Information** | | |
|---|---|---|---|---|---|---|
| From | To | R/Ω | X/Ω | Bus | P/kW | Q/kVAR |
| 1 | 2 | 0.8 | 0.9 | 1 | 8 | 3 |
| 1 | 4 | 0.9 | 1.4 | 2 | 10 | 4 |
| 2 | 3 | 0.8 | 1.0 | 3 | 21 | 5 |
| | | | | 4 | 14 | 4 |

TABLE II DG PARAMETERS

| **Symbol** | **Description** | **Value** |
|---|---|---|
| [$k_{p1}$, $k_{p2}$, $k_{p3}$, $k_{p4}$] | Active power droop coefficients | [2, 1, 0.5, 0.67] $\cdot\pi\cdot10^{-2}$ rad$\cdot$s$^{-1}\cdot$kW$^{-1}$ |
| [$k_{q1}$, $k_{q2}$, $k_{q3}$, $k_{q4}$] | Reactive power droop coefficients | [4, 2, 1, 1.3] $\cdot10^{-1}$ V$\cdot$kVAR$^{-1}$ |
| $J$ | Normalized inertia of generators | 1 kg$\cdot$m$^2$ |
| $\omega_b$ | Base frequency of generators | 377 rad/s |
| $D_p$ | Damping factor of generators | 1 N$\cdot$m |
| [$k_1$, $k_2$] | Proportional and integral gain of PI control | [5, 1] $\cdot10^5$ |
| $\sigma$ | Power output filter time constant | 0.05 s |
| $\sigma_v$ | Voltage low-pass filter time constant | 0.1 s |

### B. *Case Studies*

It should be noted that in practice the latency occurs in the communicational network is never a constant. However,

it is valuable to estimate the latency threshold from system parameters. The worst-case scenario is that every edge has the longest potential latency. In practice, the system always performs better than this worst scenario since latency is randomly distributed between zero and the longest potential latency, therefore the stability is promised. Note that time-domain circuit-based simulations using realistic DG models are utilized in this section to validate performance of the proposed framework.

*1) Model Validation: Barely Stable and Unstable Case*

Using the sample microgrid structure described in subsection IV. *A*, a time-domain simulation using *Simulink®* is carried out to test the accuracy of the proposed Nyquist-based frequency-domain analysis.

The communication network structure for this case is a directed spanning tree as shown in Fig. 5. In this case, two scenarios are simulated with the longest potential latency being 1s and 1.1s, respectively.

The Nyquist plots of these two scenarios are both shown in Fig. 6. The left side of this figure is an overview of the Nyquist contours, and the right side is the zoom-in of this plot around (-1+*j*0). Time domain simulation is carried out to validate the result of the small-signal analysis. It can be seen from Fig. 6 that the number of encirclements of (-1+*j*0) increases from 0 to 1 when the latency increases from 1.0 to 1.1s. This suggests that increasing the latency from 1.0 to 1.1s will cause the system to be unstable.

The simulation result depicted in Fig. 7 represents the time-domain response of the voltage and frequency output of each DG. Due to reactive sharing, the voltage output of each DG is different. Both the voltage and frequency waveforms are convergent when the delay is 1.0s, but fail to converge when the delay is 1.1s. Simulation results verify the prediction by the proposed frequency-domain method.

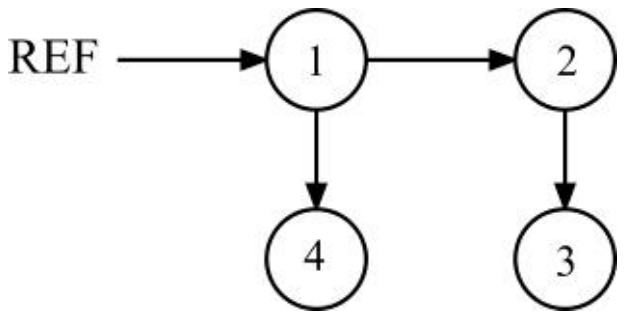

Fig. 5 Model validation: network topology

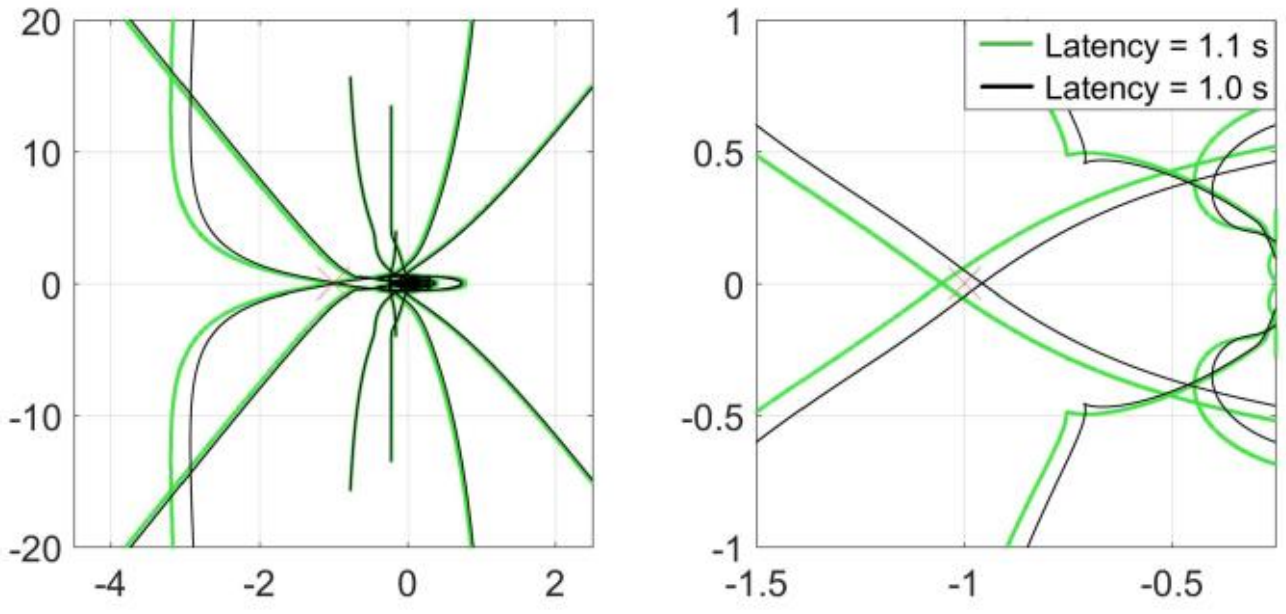

Fig. 6 Model validation: Nyquist plots

*2) Impact of Reactive Power on System Performance*

In this case, the impact of reactive power (inductive load) on the system performance is analyzed, while other system parameters and network topology are identical to case 1. As introduced in subsection III.*C*, the performance of a microgrid can be evaluated using GM and PM of the Nyquist plot, and the rate of convergence can be evaluated using GM. Again, the latency is assumed to be uniform in this case.

The impact of the reactive power level to GM and PM are shown in Fig. 8. In this case, the reactive power load at bus 1 increases from -10 to 30 kVA (the equilibrium point is also changed accordingly). Two scenarios are simulated with uniformed latency as 1.0 and 0.95s, respectively. For both scenarios, the reactive power level significantly influences the performance of the entire system, and instability may occur due to a large amount of reactive power. In contrast, negative reactive power injections can improve system performance.

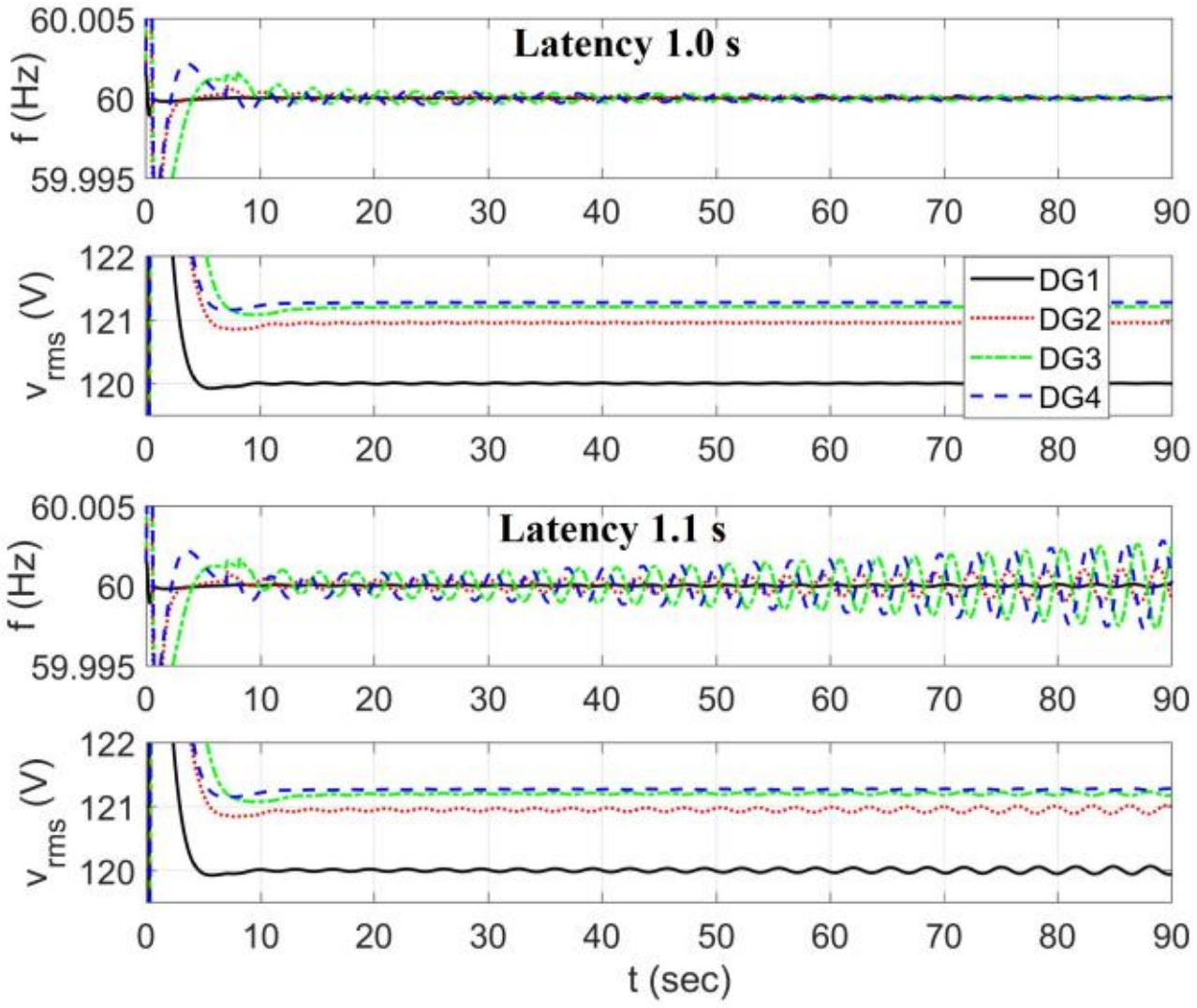

Fig. 7 Model validation: time-domain simulation

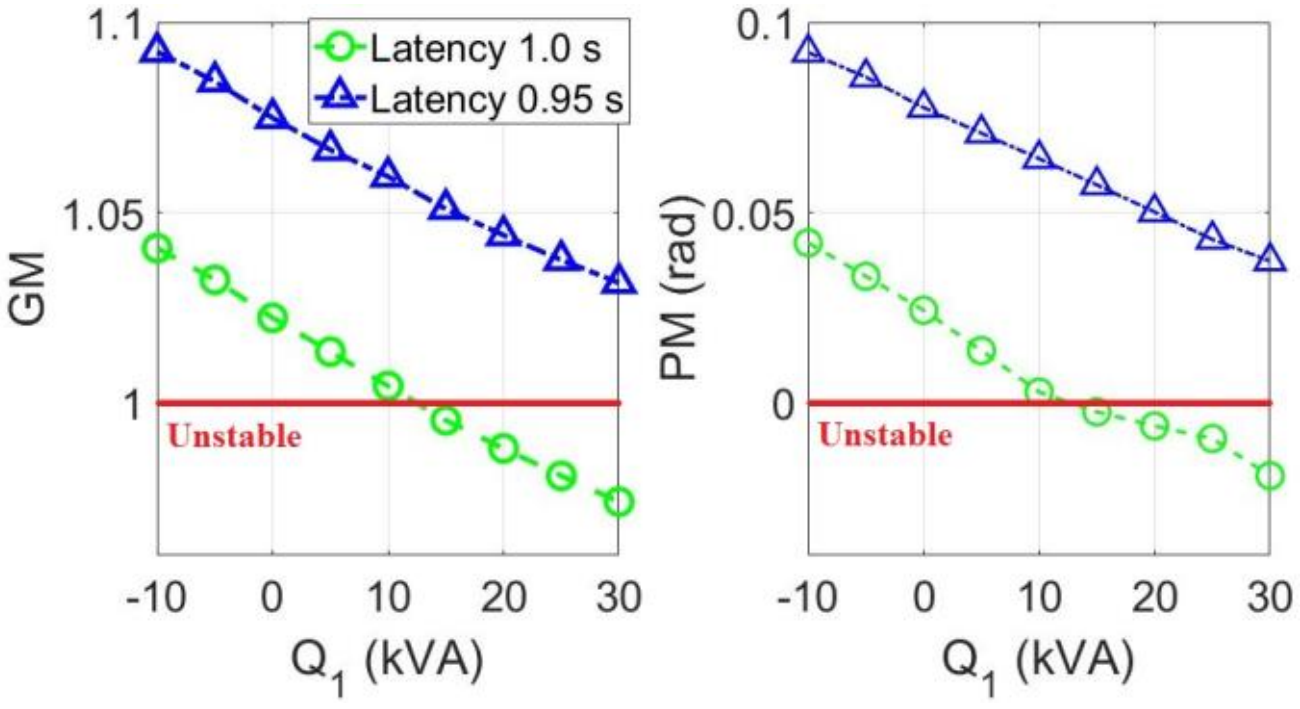

Fig. 8 Reactive power impact on system performance

*3) Impact of DG Inertia and Damping Torque Factor on System Performance*

In this case study, the impact on system performance from DG's inertia and damping factor is analyzed. The generator inertia is changed from 1 to 50 kg·m$^2$, and the damping torque factors of generators are changed from 1 to 50 N·m·s simultaneously. Other system parameters are the same as shown in Tables I and II. The communication latency is assumed uniformly 1s. Fig. 9 shows the result of this study. The red plane is the stability boundary below which the system becomes unstable. It can be seen that increasing inertia and decreasing damping factor are harmful to the system stability.

The proposed system with latency is complex and needs more discussions on overall stability with respect to inertia and damping factors. A direct reason in this situation is that, from (39), increasing inertia or decreasing damping effect will further push the response phase closer to -180 degrees around the dominant harmonic frequency (around 0.3Hz from Fig. 7).

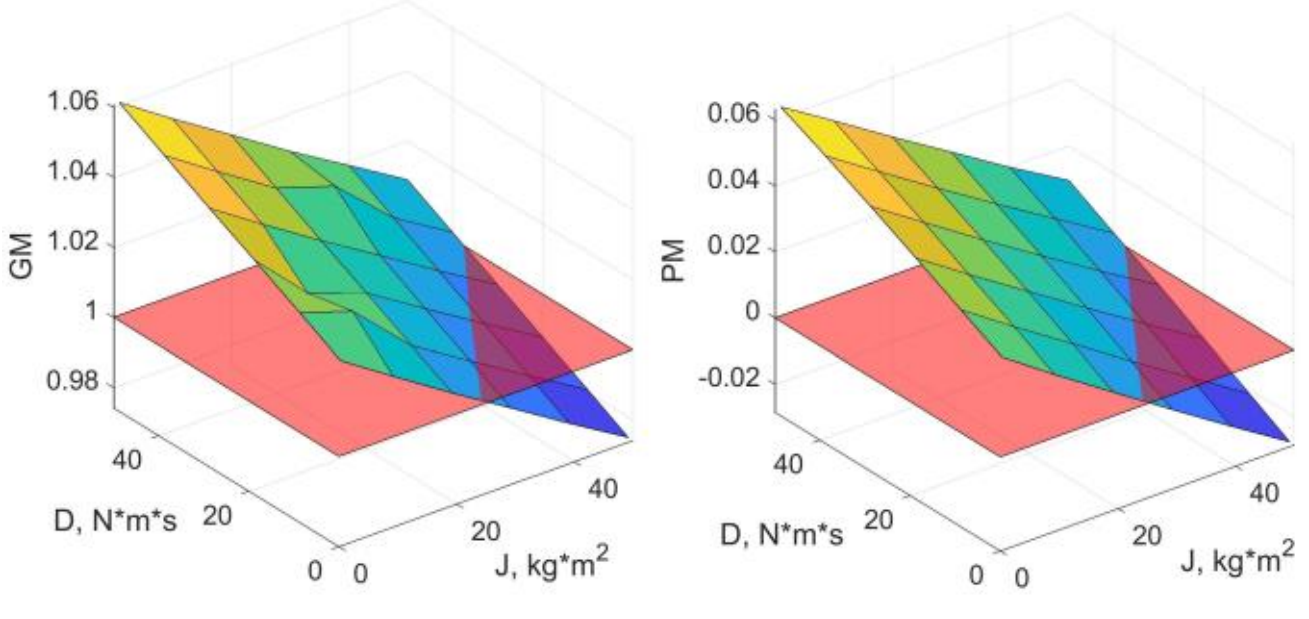

Fig. 9 DG inertia and damping impact on system performance

### 4) Impact of Network Topology on System Performance

In this case, two network topologies other than the binary tree one are simulated. The first topology is a fully-mesh network, in which every node is connected to all other nodes. The second topology is the one optimized for the uniform 1.0s latency, which is shown in Fig. 10. This can be obtained by taking an exhaustive search for the flowing optimization problem:

$$\begin{aligned} &\text{maximize} && f(\mathbf{A} = [a_{ij}]) \\ &\text{subject to} && \textstyle\sum_j a_{ij} + \theta_i > 0, \\ &&& a_{ij} = 0 \text{ or } 1, \\ &&& a_{ij} = 0 \text{ for } i = j. \end{aligned}$$

The optimization objective $f(\mathbf{A})$ is the GM or equivalently, the absolute value of the largest (negative) non-zero eigenvalue of the entire microgrid system.

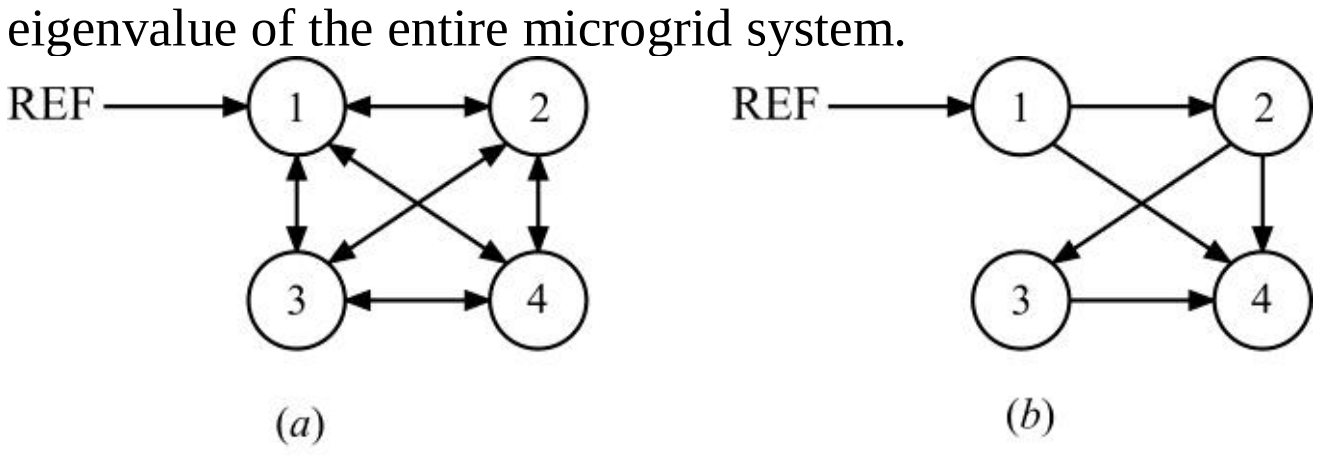

Fig. 10 (a) Full-mesh topology (b) Optimal topology for uniform 1.0 s latency.

The GM and PM versus communication latency dependences corresponding to two interested network topologies are depicted in Fig. 11. From this figure, although the fully-mesh topology results in a better robustness to the latency, its rate of convergence is inferior to the “optimal” topology for lower latencies (less than 2.1s). This phenomenon can be verified by the simulation result shown in Fig. 12 with uniform 1.0 s latency. Therefore, for different network topologies, there exists a trade-off between the rate of convergence in low latency range and the robustness to high latency range. In practice, for low latency communication infrastructures, a microgrid may improve its rate of convergence by eliminating some non-crucial communication edges.

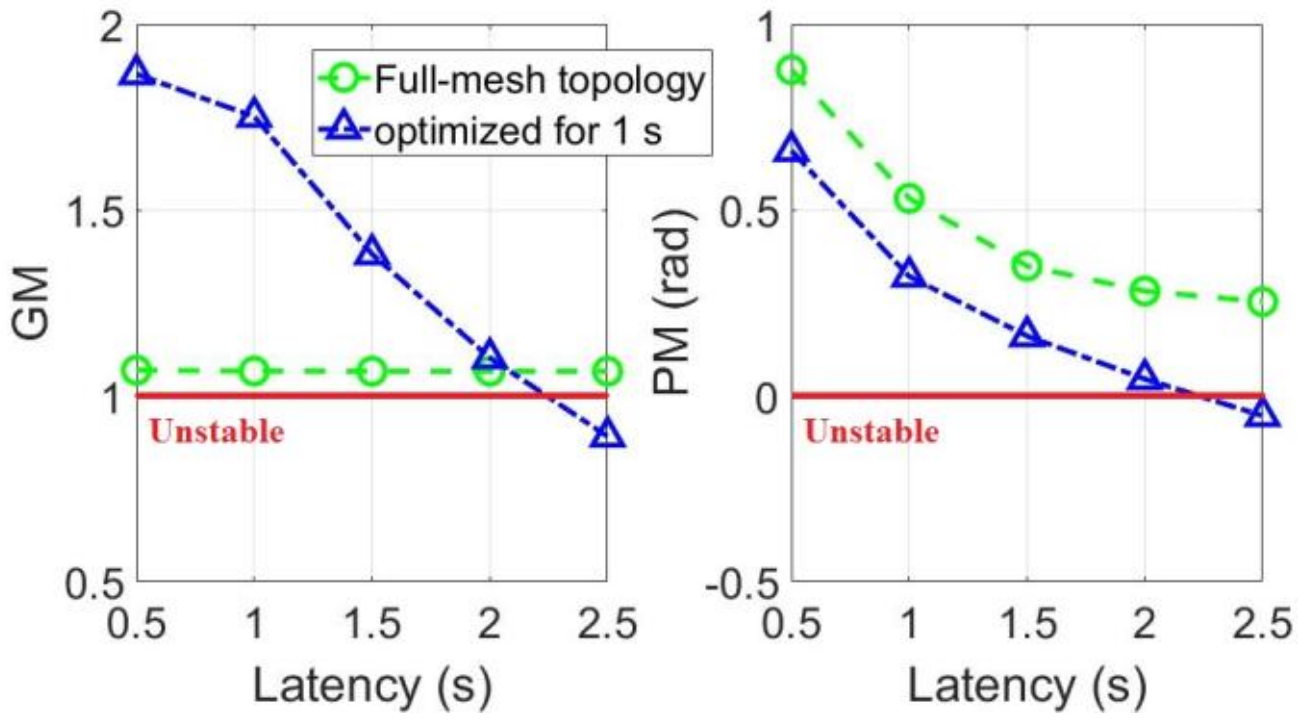

Fig. 11 Performance comparison between the full-mesh graph and the “optimal” topology

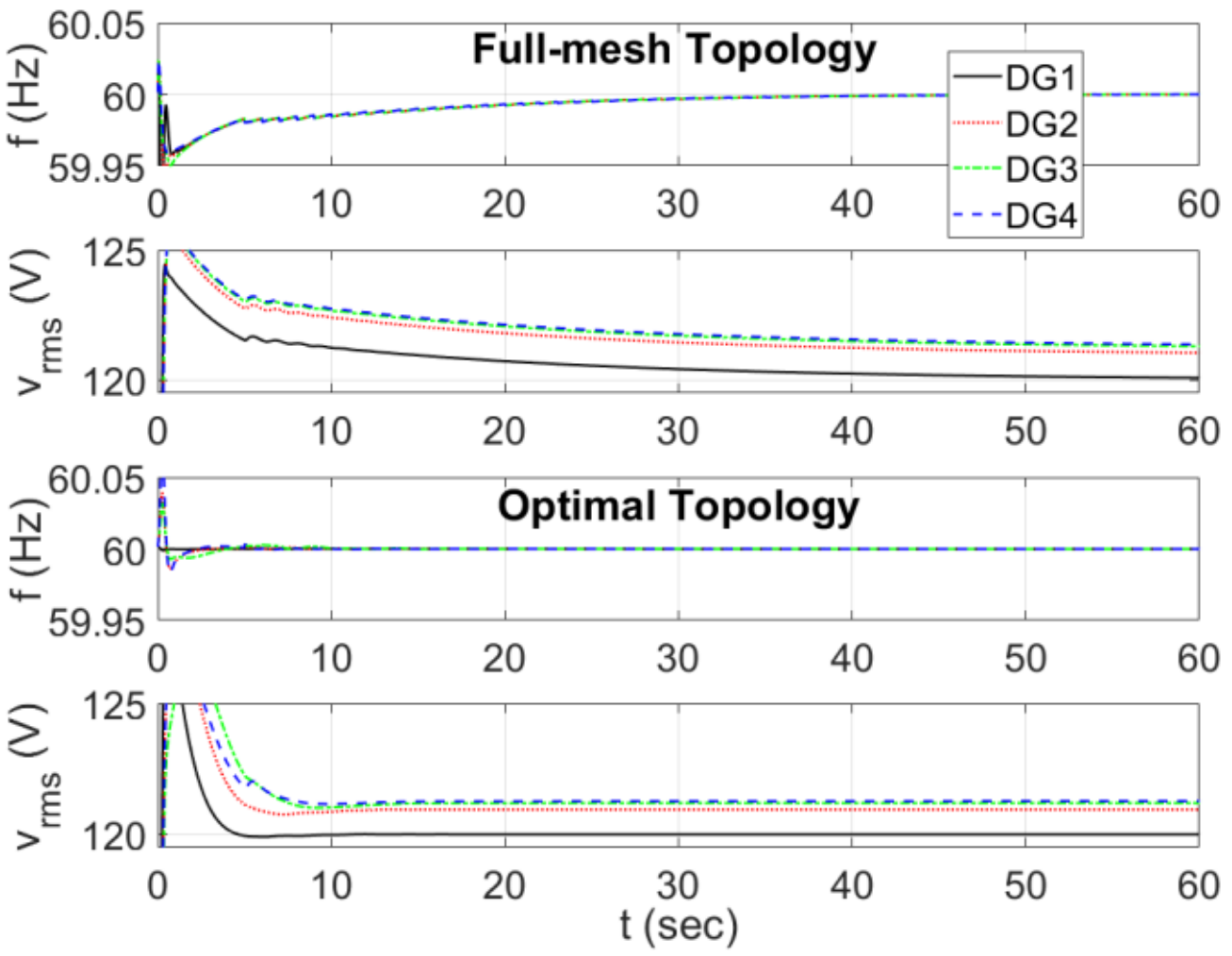

Fig. 12 Time-domain verification for the full-mesh and the “optimal” topology

### 5) Impact of Communication Stabilizer

A simple but effective approach to improve system tolerance of communicational misbehaviors is to add a stabilizer transfer function to all communication signals or data from neighbors before utilized by controllers. Compared to the techniques which compensate latencies by predictors [31, 32], this method is much easier to be practically implemented. Consider the following transfer function:

$$F(s) = \frac{0.5s^2 + s + 1}{(s+1)^2} \tag{47}$$

For different system architectures and network latencies, the numeric parameters in (47) are can be modified accordingly, which is a future research topic. This transfer function is to be embedded into the communicational path either in the sender side or the receiver side. Fig. 14 illustrates an implementation of stabilizer transfer function in the receiver frequency input, as a modification from Fig. 4(a). The same function is applied to the voltage input as in Fig. 4(b).

For the same system setup as case 1 with communication diagram shown in Fig. 5 and a uniform communication latency of 2 s, Nyquist plots and results from time-domain simulation are obtained and shown in Fig. 14 and Fig. 15, respectively. Compared to Fig. 7, which shows the system becomes unstable with 1 second communication latency, the system tolerance to latency is increased to over 2 seconds with the stabilizer.

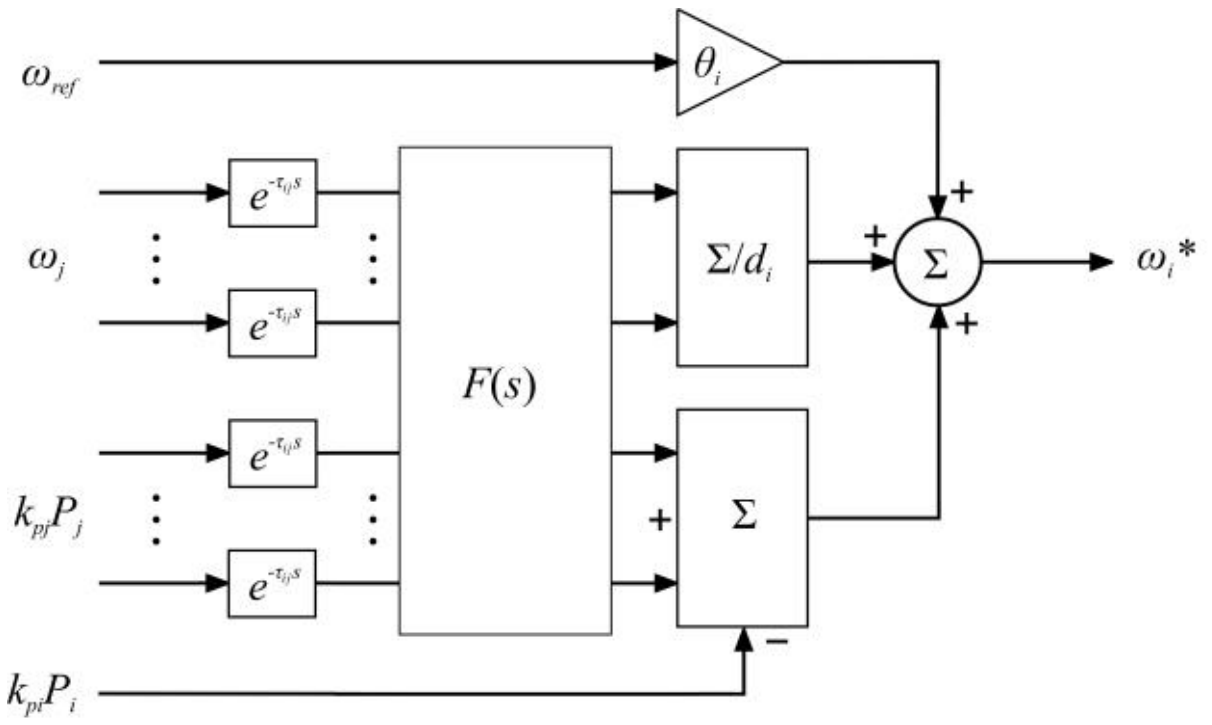

Fig. 13 Communication stabilizer for frequency input

The discussion for this case indicates that using this framework, it is convenient to add, reduce or modify system design features and to test system response in an efficient way.

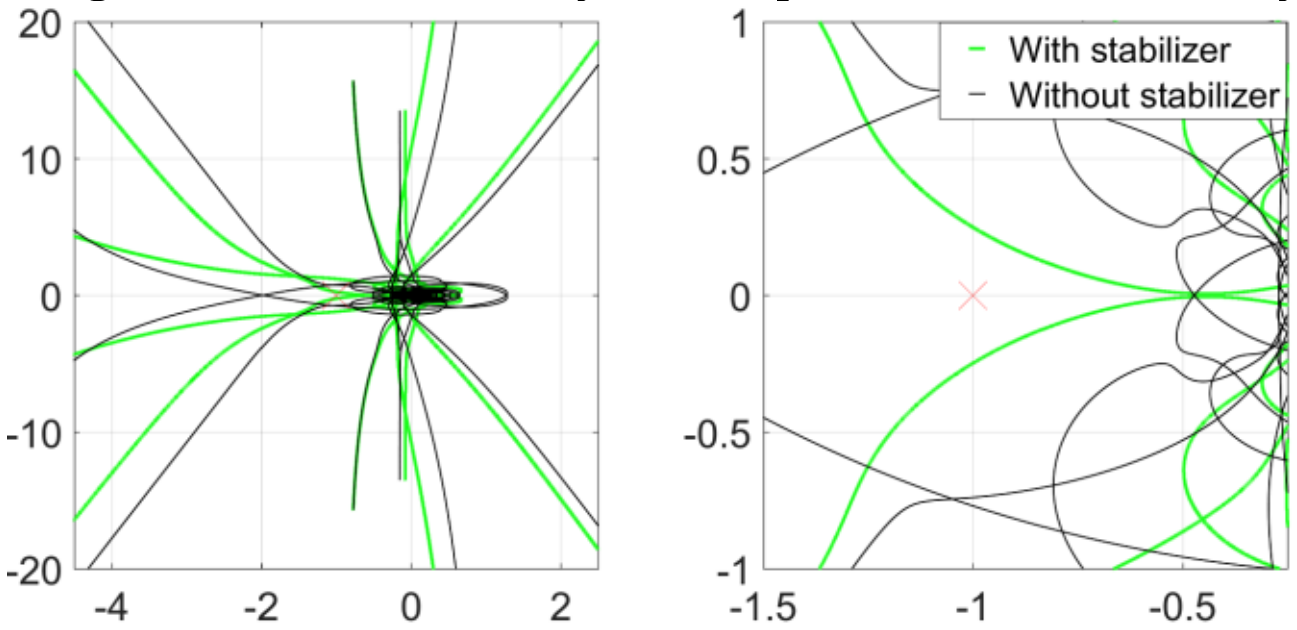

Fig. 14 System characteristic with communication stabilizer

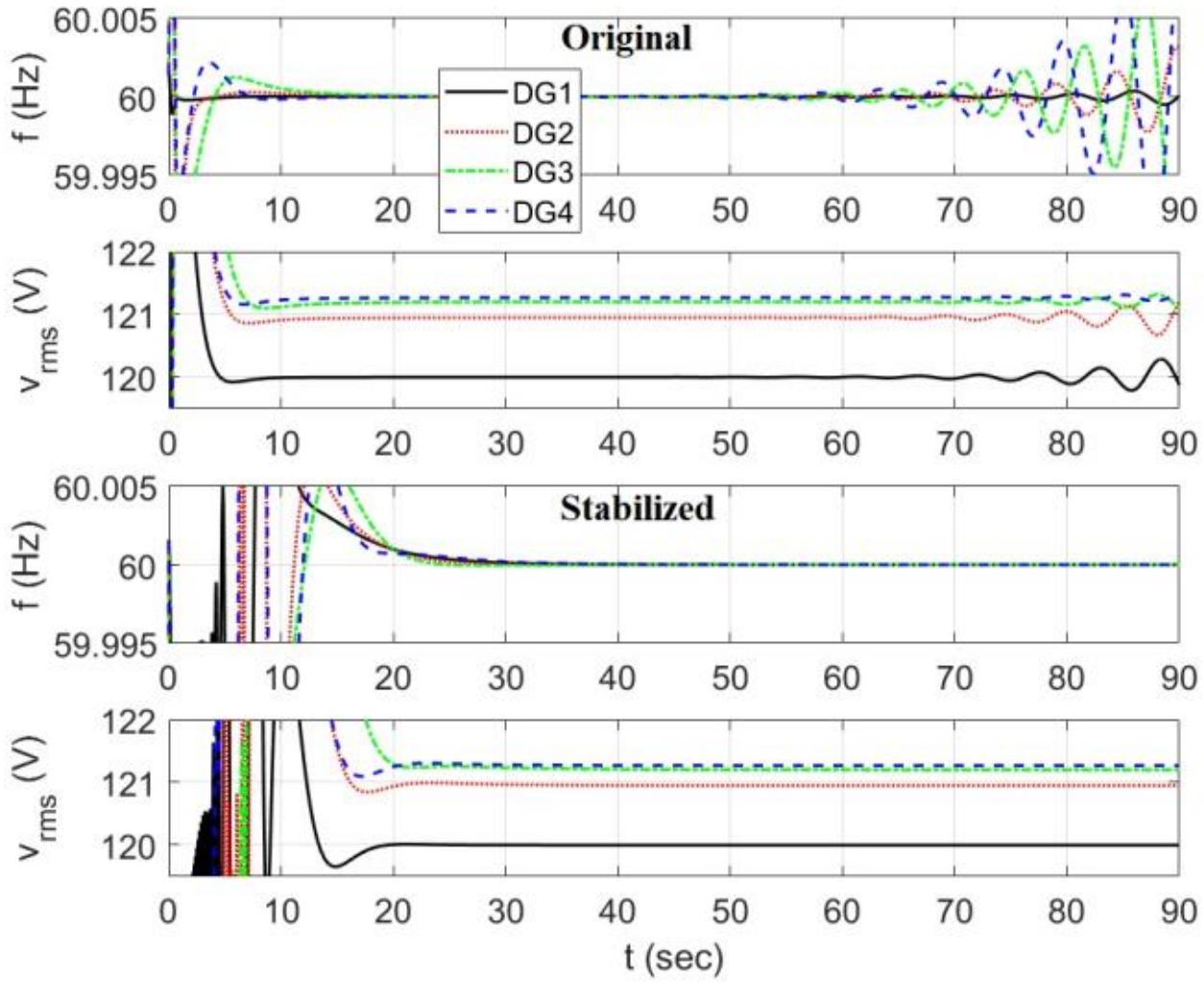

Fig. 15 Time-domain simulation with communication stabilizer

### 6) Discussion on Varying Latency

In practice, communication networks are not ideal. A number of factors affecting the communication may happen randomly and quite frequently [33]. Therefore, it is reasonable to consider a random latency value distributed within a certain range.

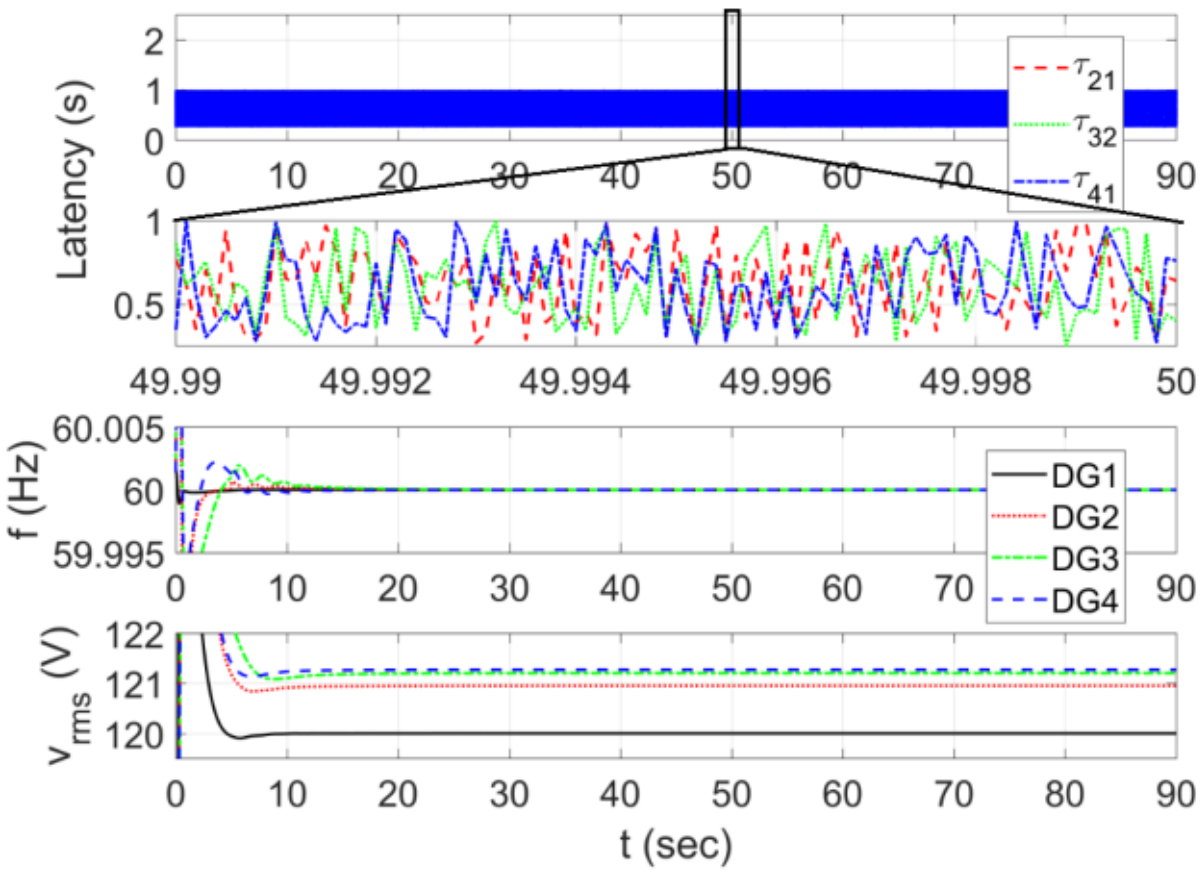

(a) random latency ranges from 0.25s to 1s

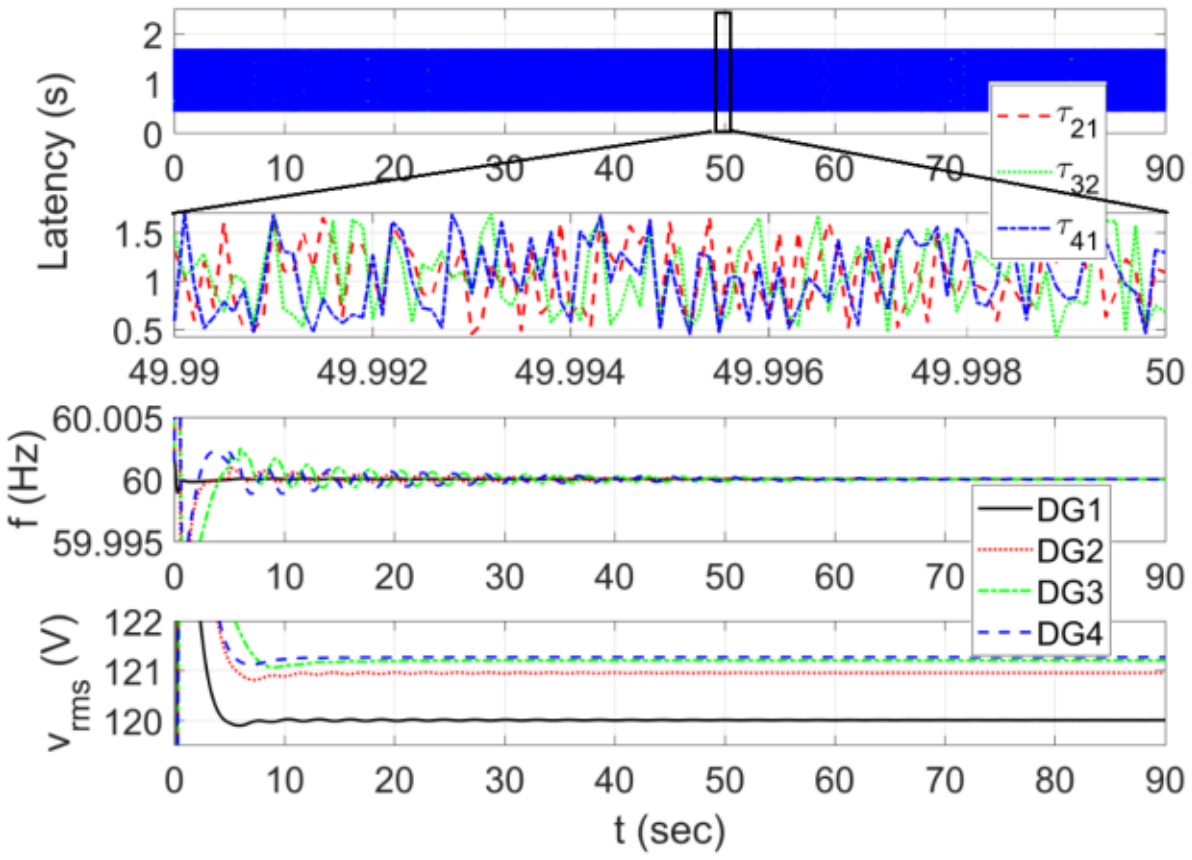

(b) random latency ranges from 0.425s to 1.7s

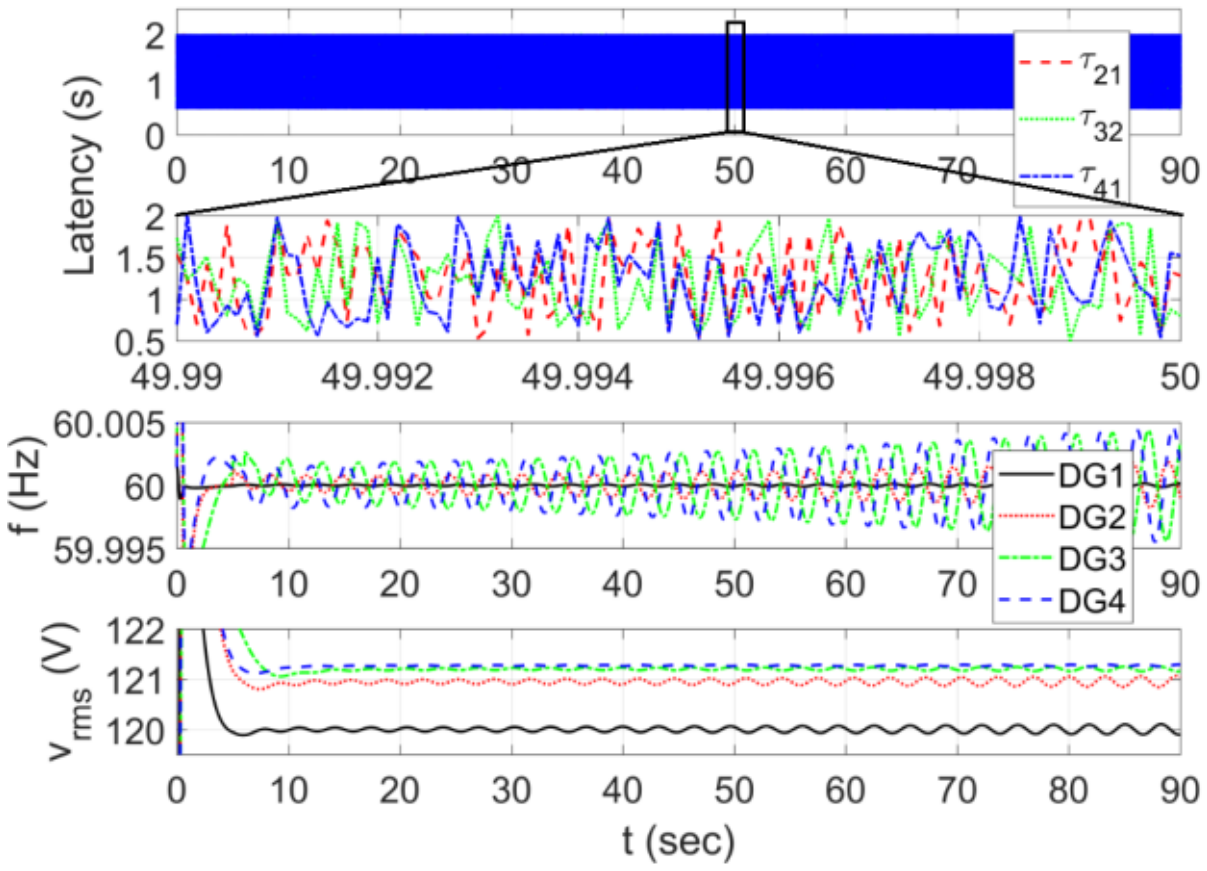

(c) random latency ranges from 0.5s to 2s

Fig. 16 Simulation results with time-varying latency

To show the relationship between the estimated constant latency and the realistic random latency, a set of simulations are carried out with results shown in Fig. 16. The system setup is the same as case I, where there are three communicational edges: $DG_1$ to $DG_2$ with latency $\tau_{21}$, $DG_2$ to $DG_3$ with latency $\tau_{32}$ and $DG_1$ to $DG_4$ with latency $\tau_{41}$. Each latency is independent and varies in a random pattern. Each random latency value is uniformly distributed from a maximum value $\tau_{max}$ to a quarter of this maximum value $\tau_{max}/4$. In Fig. 16, three cases are shown with $\tau_{max}$ valued at 1.0s, 1.7s and 2.0s, respectively. The first subplot of Fig. 16 (a) (b) (c) is the random latency during a 90-second interval, while the second subplot zooms into the latency during 49.99 - 50.00 s.

From section IV.B(1), the latency threshold of this system has been determined to be around 1.0s to 1.1s. As the simulation results show, 1) system is stable when the random latency ranges from 0.25s to 1s; 2) when the random latency ranges from 0.425s to 1.7s, with an average value around 1.06s, the system is still stable; 3) when the latency ranges from 0.5s to 2s, system becomes unstable. In conclusion, the small-signal analysis gives an appropriate estimate of the average latency values in real situations.

## V. Conclusion and discussions

This paper presents a framework for stability analysis and performance evaluation of microgrids under distributed control considering communication latency and its uncertainty. The consensus-based distributed control based on frequency-domain small-signal analysis and the generalized Nyquist theorem are implemented. The proposed approach generalizes communication latency, source and load characteristics, rotor inertia, and network topology. The effectiveness of the proposed method is verified through different case studies with comparison to time-domain simulations. This work may also be modified to adapt to other types of distributed control for power systems, which utilize other communication and control disciplines.

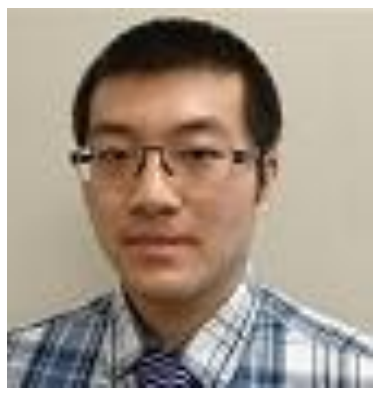

**Yimajian Yan** (S'10) received the B.E. degree in electrical engineering from Stony Brook University, NY in 2009. He is currently pursuing the Ph.D. degree in electrical and computer engineering at Purdue University, IN. His research interests include power systems, applied control, and the overlap between power electronics and signal/power integrity.

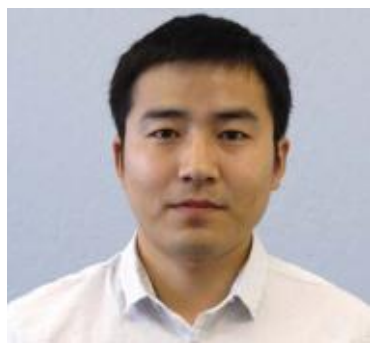

**Di Shi** (M'12-SM'17) received the B. S. degree in electrical engineering from Xi'an Jiaotong University, Xi'an, China, in 2007, and M.S. and Ph.D. degrees in electrical engineering from Arizona State University, Tempe, AZ, USA, in 2009 and 2012, respectively. He currently leads the PMU & System Analytics Group at GEIRI North America, San Jose, CA, USA. His research interests include WAMS, Energy storage systems, and renewable integration. He is an Editor of IEEE Transactions on Smart Grid.

**Bibin Huang** is currently with State Grid Energy Research Institute as a researcher. His research interest includes power system planning, and grid integration of renewables.

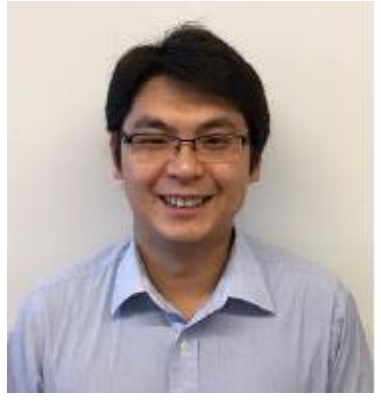

**Desong Bian** (S'12 - M'17) received his B.S. degree from Department of Electrical and Computer Engineering, Tongji University, Shanghai, China in 2007, M.S. degree from Department of Electrical and Computer Engineering, University of Florida, Gainesville, FL, USA in 2011, and Ph.D. from the School of Electrical and Computer Engineering, Virginia Tech, Arlington, VA, USA in 2016. He is currently an Engineer with GEIRI North America, San Jose, CA, USA. His research interests include PMU related applications, demand response, communication network for smart grid, etc.

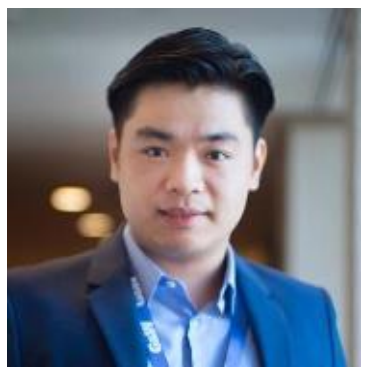

**Zhehan Yi** (S'13-M'17) received the B.S. in electrical engineering from Beijing Jiaotong University, Beijing, China in 2012, and the M.S. and Ph.D. in electrical engineering from The George Washington University, Washington, DC, USA, in 2014 and 2017, respectively. He is currently a Power System Research Engineer with GEIRI North America, San Jose, CA, USA. His research interests are power system control, microgrids, and renewable integration.

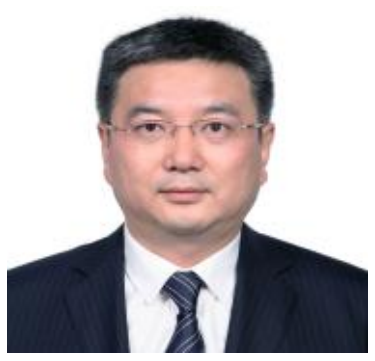

**Zhiwei Wang** (M'16-SM'18) received the B.S. and M.S. degrees in electrical engineering from Southeast University, Nanjing, China, in 1988 and 1991, respectively. He is President of GEIRI North America, San Jose, CA, USA. Prior to this assignment, he served as President of State Grid US Representative Office, New York City, from 2013 to 2015, and President of State Grid Wuxi Electric Power Supply Company from 2012-2013. His research interests include power system operation and control, relay protection, power system planning, and WAMS.